\documentclass[
twocolumn, preprintnumbers, amsmath,amssymb]{revtex4}

\usepackage{graphicx}
\usepackage{dcolumn}
\usepackage{bm}

\newcommand\ee{\end{equation}}
\newcommand\be{\begin{equation}}
\newcommand\eea{\end{eqnarray}}
\newcommand\bea{\begin{eqnarray}}
\newcommand{\sfrac}[2]{{\textstyle\frac{#1}{#2}}}
\newcommand\di{\partial}

\begin{document}


\title{Galilean currents and charges}

\author{Alberto Nicolis}
\email{nicolis@phys.columbia.edu}

\affiliation{%
Physics Department and Institute for Strings, Cosmology, and Astroparticle Physics,\\
Columbia University, New York, NY 10027, USA
}%

\date{\today}

\begin{abstract}
We derive the Noether currents and charges associated with an internal galilean invariance $\pi(x) \to \pi(x) + b_\mu x^\mu$---a symmetry recently postulated in the context of so-called galileon theories. Along the way we clarify the physical interpretation of the Noether charges associated with ordinary Galileo- and Lorentz-boosts.
\end{abstract}

\maketitle


\noindent
There has been recent interest in galileon theories. In their simplest version  \cite{NRT}, these correspond to an effective field theory for  a Goldstone boson $\pi$ that is invariant under {\em internal} Galilean transformations 
\be \label{galilean_pi}
\pi(x) \to \pi(x) + b_\mu x^\mu \; .
\ee 
This theory has a number of interesting and novel field-theoretical properties, at the classical level as well as at the quantum-mechanical one \cite{NR, NRT, EHHNW, CN}, which make it potentially relevant for IR-modifications of gravity \cite{NRT}, for consistent violations of the null energy condition within QFT \cite{NRT2}, and for alternatives to slow-roll inflation \cite{CNT, BdRST, CDMNT}.
Several generalizations of the minimal galileon have been proposed, and some of them will be briefly touched upon in the following.

Here we will derive the conserved local currents and global charges that the Noether theorem associates with the symmetry (\ref{galilean_pi}).
However, it is instructive to consider first the conservation laws associated with {\em ordinary} Galilean invariance, that is with the symmetry
\be
\vec x \to \vec x + \vec v_0 \, t \; , \label{galilean_x}
\ee
because their interpretation presents some subtleties.
We will briefly discuss the analogous case of  Lorentz invariance below.
The position $\vec x$ can be a dynamical degree of freedom, like for a non-relativistic mechanical system made up of point particles and parameterized by their positions, or an integration variable, that is an argument for fields in a field theory.  Typically systems that are galilean invariant are also translationally invariant---for instance this is guaranteed if the system is galilean invariant and {\em time}-translationally invariant---so that we also have a symmetry
\be
\vec x \to \vec x + \vec  x_0 \; . \label{shift_x}
\ee
Now we apply the standard derivation of the Noether theorem. Let us start with the symmetry (\ref{shift_x}). Because of it, the variation of the action under an infinitesimal weakly time-dependent translation with parameter $\vec x_0(t)$ must start at order $\dot {\vec x}_0$:
\be \label{deltaS_x}
\delta S \simeq \int \! dt \, \dot {\vec x}_0 \cdot \vec P \; . 
\ee
On a solution to the equations of motion the action is stationary, which implies the conservation of $\vec P$.
As is well known, the charge $\vec P$ associated with spatial translations is the total momentum of the system.

Now, the fact that the system is also invariant under (\ref{galilean_x}) implies that, in fact, $\delta S$ in (\ref{deltaS_x}) should start at order $\ddot {\vec x}_0$,
\be \label{deltaS_x_2}
\delta S \simeq - \int \! dt \, \ddot {\vec x}_0 \cdot \vec \Xi
\ee
(the minus sign upfront is for notational convenience).
This is equivalent to saying that $\vec P$ in (\ref{deltaS_x}) is itself a total time derivative,
\be \label{PdotQ}
\vec P = \dot {\vec \Xi} \; ,
\ee
which combined with the conservation of $\vec P$, implies that on all solutions $\vec \Xi$ is a linear function of time:
\be \label{linear_Q}
\vec \Xi(t) = \vec \Xi_0 + \vec P \,  t \; .
\ee
$\vec \Xi(t)$ is nothing but the center of mass position $\vec X_{\rm cm}$ times the total mass of the system. Eq.~(\ref{linear_Q}) is the global conservation law associated with Galilean invariance.
To explicitly check this, let's perform a galilean transformation (\ref{galilean_x}) with mildly time-dependent $\vec v_0(t)$. We have
\be \label{deltaS_v} 
\delta S \simeq \int \! dt \, \dot {\vec v}_0 \cdot \vec {\cal Q}  \; ,
\ee
implying conservation of the quantity $\vec {\cal Q}$ on  a solution. We can express $\vec {\cal Q}$ in terms of $\vec P$  and $\vec \Xi$, by noticing that the time-dependent Galilean transformation we are performing can also be viewed as a time dependent translation with parameter $\vec x_0(t) = \vec v_0(t) t$. From (\ref{deltaS_x_2}) and (\ref{PdotQ}) we have
\be
\delta S \simeq  \int \! dt \, \big(\dot {\vec v}_0 t + \vec v_0 \big) \cdot \dot {\vec \Xi} 
\ee
which compared with (\ref{deltaS_v}) yields
\be
\vec {\cal Q} = \vec P \, t - \vec  \Xi \; .
\ee
The conservation of $\vec {\cal Q}$ is thus equivalent to eq.~(\ref{linear_Q}), with $\vec \Xi_0 = -\vec {\cal Q}$. Therefore, we see that the conservation law associated with Galilean invariance is more conveniently rephrased as the statement that there is a quantity---the center of mass position---that evolves linearly in time. This is in addition to, and not a consequence of, the conservation of the total momentum. Of course if $\vec P$ is conserved, we can always say that $\vec P \, t$ evolves linearly in time. But eq.~(\ref{linear_Q}), with $\vec \Xi$ defined in (\ref{deltaS_x_2}), is a much stronger statement. It says that there is a quantity $\vec \Xi(t)$ that is {\em a local functional of the dynamical variables, with no explicit time dependence}, that happens to evolve linearly in time. It is as strong as an ordinary conservation law, whereby a local functional of the dynamical variables happens to be constant in time. To be completely clear, by `local functional' we mean, in the point-particle mechanical case, a function of the particles' coordinates and their time derivatives, all evaluated at the same time, and in the field theory case, the space integral of a local density. Indeed, the standard expressions for $\vec \Xi$ for Galilean mechanics
is
\be
\vec \Xi_{\rm mech.} (t)  =  \sum_a m_a \vec x_a(t) \\
\ee

To convince ourselves that the linear evolution of $\vec \Xi$ and the conservation of $\vec P$ are really independent physical laws, we can consider a mechanical system that is invariant under translations but not under galilean boosts, like for instance one made up of many point-particles interacting via a two-body potential and with non-Newtonian kinetic energies:
\be
S = \int \! dt  \,  \sum_a m_a \big( \sfrac12 \dot {\vec x}_a ^{\, 2} +  \alpha \,  \dot {\vec x}_a ^{\, 4}  \big) -\sum_{a<b}  V_{ab}(|\vec x_a - \vec x_b|) \; .
\ee
The system is invariant under translations, and thus the total momentum
\be
\vec P = \sum_a m_a  \dot {\vec x}_a \big( 1 + 4 \alpha \, \dot {\vec x}_a^{\, 2}  \big)
\ee
is conserved. Nevertheless, for nonzero $\alpha$ the system is not invariant under galilean transformations, and as a consequence $\vec P$ is not associated with the time derivative of a collective coordinate---there is no local combination of $\vec x_a$ and $\dot {\vec x}_a$ that evolves linearly in time. {\em The conservation of $\vec P$ does not imply that the system as a whole moves at constant speed}. It only does for Galilean (or Lorentz-) invariant theories.

Of course all of the above discussion can be straightforwardly generalized to the relativistic case.
Indeed for a Lorentz-invariant field theory the Noether charges associated with the Lorentz boosts are (see e.g.~\cite{weinberg})
\be
J^{0i} = t P^i - \int \! d^3 x \, T^{00} x^i \; ,
\ee
where $T^{\mu\nu}$ is the (symmetric) stress-energy tensor.
They are close relatives of the angular momentum, but unlike it,  they are usually glossed over. From the above discussion their physical interpretation is clear: their conservation implies the existence of a collective coordinate
\be
\vec \Xi(t) = \int \! d^3 x \, T^{00} (x) \, \vec x \equiv M \vec X_{\rm cm} (t)
\ee
 that evolves linearly with time.

We can now move on to the galileon case, and run an analogous derivation. The theory enjoys an internal Galilean invariance (\ref{galilean_pi}) as well as a more conventional shift invariance
\be \label{shift_pi}
\pi(x) \to \pi(x) + c \; .
\ee
The latter yields a standard Noether current $j^\mu$, which we get by performing a weakly $x$-dependent infinitesimal shift:
\be \label{deltaS_pi}
\delta S \simeq \int \! d^4 x \, \di_\mu c \,  j^\mu \; .
\ee
However, because of the symmetry (\ref{galilean_pi}), this variation should in fact start at second order in derivatives of $c(x)$:
\be \label{deltaS_pi_2}
\delta S \simeq - \int \! d^4 x \, \di_\mu \di_\nu c \,  \xi^{\mu\nu} \; ,
\ee
so that the current $j^\mu$ is a total divergence:
\be \label{jdJ}
j^\nu = \di_\mu \xi^{\mu\nu} \; .
\ee
(From now on, without loss of generality, we take $\xi^{\mu\nu}$ to be symmetric.)
The global charge associated with $j^\mu$ is of course
\be \label{q}
Q = \int \! d^3 x \, j^0 \; .
\ee
Now, the $\pi$ equation of motion is equivalent to the conservation of $j^\mu$,
\be
\di_\mu j^\mu = \di_\mu \di_\nu \xi^{\mu\nu} = 0
\ee
which implies that the spacial integral of $\xi^{00}$ evolves linearly in time---it is the analogue of our center-of-mass coordinate above:
\bea
\di_0^2 \int \! d^3 x \, \xi^{00} = 2 \di_0 \int \! d^3 x \, \di_i \xi^{0i} +  \int \! d^3 x \, \di_i \di_j \xi^{ij} = 0
\eea
Moreover, its time-derivative is the shift Noether charge $Q$---the analogue of the total momentum: 
\be
\di_0 \int \! d^3 x \, \xi^{00} = \int \! d^3 x \, \big( \di_\mu \xi^{0\mu} - \di_i \xi^{0i}  \big) = \int \! d^3 x \, j^0
\ee
In conclusion, there is a local functional of the fields
\be
\Xi (t)\equiv \int \! d^3 x \, \xi^{00}(x)
\ee
that on all solutions happens to evolve linearly in time,
\be
\Xi (t) = \Xi_ 0 + Q \,  t \; . 
\ee
Like in our original example, this statement can be seen as a consequence of the internal galilean invariance (\ref{galilean_pi}), but there is more. After all, our galilean symmetry has four parameters, and here we discovered just one global conservation law. There should be four locally conserved currents, and four associated global charges. To identify them, we run Noether's theorem with a weakly $x$-dependent galilean shift (\ref{galilean_pi}), with parameter $b_\mu(x)$. The variation of the action starts at order $\di b$:
\be \label{deltaS_db}
\delta S \simeq \int \! d^4 x \, \di_\mu b_\alpha \, {\cal J}^{\mu \, (\alpha)}  \; ;
\ee
$\alpha$ labels the symmetry, and to avoid confusion for the moment we use a different notation than for Lorentz indices. We have four conserved  currents
$ {\cal J}^{\mu \, (\alpha)} $, which we can relate to $j^\mu$ and $\xi^{\mu\nu}$ using the same trick as above. We think of our $x$-dependent galilean transformation  as an  $x$-dependent shift (\ref{shift_pi}), with parameter $c(x) = b_\mu(x) x^\mu$. By equating (\ref{deltaS_db}) with (\ref{deltaS_pi_2}) we thus get
\be \label{calJ}
{\cal J}^{\mu \, (\alpha)} = \xi^{\mu\alpha} - x^\alpha j^\mu \; .
\ee
These are the four Noether currents associated with the symmetry (\ref{galilean_pi}). They are conserved on the eom,
\be
\di_\mu {\cal J}^{\mu \, (\alpha)} = \di_\mu \xi^{\mu\alpha} - \di_\mu \big( x^\alpha j^\mu \big) = - x^\alpha \di_\mu j^\mu \; ,
\ee
because $j^\mu$ is.
The associated global charges are
\be
{\cal Q}^\alpha = \int \! d^3 x \, \xi^{0\alpha} - x^\alpha j^0 \; , 
\ee
or more explicitly:
\bea 
{\cal Q}^0 & = &  \int \! d^3 x \, \xi^{00} - t \, Q = \Xi (t) - t \, Q \label{Q0} \\
{\cal Q}^i & = & \int \! d^3 x \, \big(\xi^{0i} - x^i j^0 \big) \label{Qvec} 
\eea
The conservation of ${\cal Q}^0$ is equivalent to the linear evolution of $\Xi(t)$ we discovered above. The conservation 
of $\vec {\cal Q}$ is a more traditional conservation law, in that it does not involve an explicit time dependence.
Notice that the second piece in $\vec {\cal Q}$ is the total charge dipole of the system.

In summary, we have five locally conserved currents $j^\mu$ and  ${\cal J}^{\mu \, (\alpha)}$, and five corresponding global charges $Q$  and ${\cal Q}^\alpha$.
As an almost trivial example, we can consider the simplest system with internal galilean invariance---the free massless scalar:
\be
S = -\int \! d^4 x \, \sfrac12 (\di \pi)^2
\ee
The various local quantities we have defined above are
\bea
&&j^\mu = - \di^\mu \pi \; , \qquad \xi^{\mu\nu} = - \eta^{\mu\nu} \, \pi \; , \label{L2_1} \\
&&{\cal J}^{\mu \, (\alpha)} = - \eta^{\mu\alpha} \, \pi + x^\alpha \di^\mu \pi \label{L2_2} \; ,
\eea
which yield the conserved  charges
\bea
Q & = & {\textstyle \int} d^3 x \,  \dot \pi \\
{\cal Q}^0 & = &  {\textstyle \int}  d^3 x \, \pi   - t \, q\\
{\cal \vec  Q} & = & - {\textstyle \int}  d^3 x \, \dot \pi \vec x
\eea
The first is just the usual charge associated with shift invariance. The third is the total dipole moment of that charge. The fact that it is conserved is here a trivial consequence of the equation of motion (like all conservation laws, to some extent), but it is nonetheless a non-trivial statement (unheard of, at least). And so is the conservation of ${\cal Q}^0$, which implies that the space integral of $\pi$ grows linearly in time.

We can now derive explicit expressions for the currents we have defined, in the case of a generic galileon Lagrangian. The Lagrangian is a function of first and second derivatives of $\pi$,
\be
{\cal L} = {\cal L} (\di\pi , \di \di \pi) \; ,
\ee
with suitable Lorentz contractions to ensure galilean invariance \cite{NRT}. The shift current $j^\mu$ is readily determined, by noticing that under an infinitesimal shift (\ref{shift_pi}) with $x$-dependent parameter $c(x)$, we have
\bea
\delta S & = & \int \! d^4 x \, \frac{\di {\cal L}}{\di(\di_\mu \pi)} \di_\mu c +\frac{\di {\cal L}}{\di(\di_\mu \di_\nu\pi)} \di_\mu \di_\nu c
\eea
which compared with (\ref{deltaS_pi}) yields
\be \label{jmu}
j^\mu = \frac{\di {\cal L}}{\di(\di_\mu \pi)}  -  \di_\nu \frac{\di {\cal L}}{\di(\di_\mu \di_\nu\pi)} \; .
\ee
To compute the currents associated with galilean shifts, eq.~(\ref{calJ}), we need first to determine $\xi^{\mu\nu}$, which is defined simply as a symmetric tensor with divergence $j^\mu$---eq.~(\ref{jdJ}). The second piece in (\ref{jmu}) is manifestly the divergence of a symmetric tensor.
To rewrite the first piece also as a total divergence requires more work. The reason is that the Lagrangian is {\em not} invariant under galilean shifts (\ref{galilean_pi})---only the action is. That is, the Lagrangian is invariant only up to a total derivative, which means that
the variation of the action under an infinitesimal $x$-dependent shift 
is not manifestly of the form (\ref{deltaS_pi_2}). To proceed, we need the explicit expression for the galilean invariants. At $(n+1)$-st order in $\pi$ we have \cite{NRT}
\be \label{Ln+1}
{\cal L} _{n+1}= T^{\mu_1\nu_1 \mu_2 \nu_2 \dots \mu_n \nu_n} \, \di_{\mu_1} \pi \di_{\nu_1} \pi \, 
\di_{\mu_2} \di_{\nu_2} \pi \dots \di_{\mu_n} \di_{\nu_n} \pi
\ee
where $T$ is a tensor whose explicit form we will not need. Suffice it to say that it is 
symmetric under exchanging any two $(\mu\nu)$ pairs, and
antisymmetric under exchanging any two like indices (e.g., of the $\nu$ type) belonging to different  $(\mu\nu)$ pairs.
The latter symmetry ensures that the derivative of ${\cal L} _{n+1}$ w.r.t.~$\di \pi$ is a total divergence:
\bea
\frac{\di {\cal L} _{n+1}}{\di (\di_\mu \pi)} 
& = & \di_{\nu_2} \, T^{\mu \nu \mu_2 \nu_2 \dots \mu_n \nu_n} \, \di_{\nu} \pi \, 
 \di_{\mu_2} \pi \dots \di_{\mu_n} \di_{\nu_n} \pi  + \nonumber \\
 &&  \di_{\mu_2} \, T^{\nu \mu \mu_2 \nu_2 \dots \mu_n \nu_n} \, \di_{\nu} \pi \, 
 \di_{\nu_2} \pi \dots \di_{\mu_n} \di_{\nu_n} \pi  \label{dLdpi}
\eea
as predicted.
We can simplify this expression, by relating it to the derivative of ${\cal L}_{n+1}$ w.r.t.~$\di\di \pi$. By swapping $\nu$ with $\nu_2$ in the first term's $T$ and with $\mu_2$ in the second term's $T$, and by using the symmetry of $T$ under exchanging whole $(\mu\nu)$ pairs, we get
\be
\frac{\di {\cal L} _{n+1}}{\di (\di_\mu \pi)} =
 -\frac{2}{(n-1)} \, \di_{\nu} \frac{\di {\cal L} _{n+1}}{\di (\di_\mu \di_\nu \pi)} \; .
\ee
Plugging this into eq~(\ref{jmu}) we thus get the full $n$-th order contributions to $\xi^{\mu\nu}$:
\be
\xi^{\mu\nu}_{n+1} = -\frac{n+1}{n-1} \, \frac{\di {\cal L} _{n+1}}{\di (\di_\mu \di_\nu \pi)} \; .
\ee
Notice however that this formula only holds when the galilean invariants ${\cal L}_{n+1}$ are written as in eq.~(\ref{Ln+1}), with $T$ obeying the aforementioned symmetry properties. 
This is not the case for all the invariants explicitly displayed in ref.~\cite{NRT}, where  an integration by parts was performed on ${\cal L}_3$ to rewrite it in a more compact form, thus effectively reshuffling its dependence on $\di \pi$ with that on $\di\di \pi$. 
The `canonical' form for all the invariants relevant in 4D can be found instead in ref.~\cite{HTW}
\footnote{
Alternatively, one can use the invariants of \cite{NRT} with the replacement
${\cal L}_3 \to \sfrac13 \big( \di_\mu \pi \di_\nu \pi \, \di^\mu \di^\nu \pi - (\di \pi)^2 \Box \pi \big) $,
where we kept the same normalization as in \cite{NRT}.}.
With this qualification in mind, the currents associated with internal galilean invariance therefore  are
\be
{\cal J}_{n+1}^{\mu \, (\alpha)} =\big(x^\alpha \di_\nu -\sfrac{n+1}{n-1} \delta^\alpha_\nu \big) \frac{\di {\cal L} _{n+1}}{\di (\di_\mu \di_\nu \pi)} - x^\alpha \frac{\di {\cal L} _{n+1}}{\di (\di_\mu \pi)} \; .
\ee 
This formula cannot be applied to the lowest-order invariants, ${\cal L}_1 = \pi$ and ${\cal L}_2 = - \sfrac12(\di \pi)^2$---in deriving it we have been assuming that ${\cal L}_{n+1}$ depends non-trivially both on $\di \pi$ and on $\di\di \pi$, and that it does not depend on $\pi$. For ${\cal L}_2$, we already gave the relevant expressions in (\ref{L2_1}, \ref{L2_2}). The situation in trickier for ${\cal L}_1$. By applying Noether's theorem to it and using the identities $1 = \sfrac14 \di_\mu x^\mu$ and $x^\mu = \sfrac12 \di^\mu x^2$, we discover that its contributions to the shift current  and to the galilean one are
\be
j_1^\mu = -\sfrac14 x^\mu , \qquad {\cal J}_{1}^{\mu \, (\alpha)} = -\sfrac12 \eta^{\mu\alpha} x^2 \; .
\ee
The global charges (\ref{q}, \ref{Q0}, \ref{Qvec}) thus acquire extra pieces explicitly proportional to $t$ and to $t^2$. However, in the presence of the tadpole ${\cal L}_1 = \pi$ any solution will have non-trivial boundary conditions at spacial infinity. Like in the case of spontaneous symmetry breaking, this will generically yield divergent global charges, thus making their conservation useless. 
On the other hand the local current conservation will still be perfectly valid. For certain particularly symmetric solution, like the deSitter ones discussed in \cite{NRT}, the dynamics of perturbations about such asymptotically non-trivial solutions will still be described by a galileon theory, this time without the tadpole of course. In such a case the global charges  we derived can be used directly for the perturbations---provided one uses the perturbations' Lagrangian in our formuale.

The generalization of our results to multi-galileon theories \cite{DDEF, PSZ, HTW} should be straightforward.
Perhaps more interesting is the generalization to the so called IR-completions of the galileon, i.e.~to theories that reduce to the galileon in some appropriate limit---typically at small distances and at small field values---and that away from that limit are invariant under a different symmetry group \cite{NRT}. So far two possibilities have been proposed---promoting the galileon symmetry group (spacetime Poincar\'e plus internal shifts plus internal galilean transformations) to the conformal group $SO(4,2)$ or to the five-dimensional Poincar\'e group $ISO(4,1)$ \cite{NRT, dRT}. 
The generalization of our results to the latter case should be straightforward---as was generalizing the analogous statements we have for {\em ordinary} Galilean invariance to the Lorentz-invariant case.
The reason is that in all these cases
the transformation is linear in the relevant coordinates. As clear from our derivation, this is the crucial ingredient for our results. For instance, 
it never really mattered whether our symmetries acted linearly or non-linearly on the dynamical variables (point particle positions or fields)---the existence of a local functional that grows linearly in time follows purely from the existence of a symmetry that is linear in time.
The conformal group case is more complicated.  There, the $\pi$ shift and galilean transformations get promoted to non-linearly realized dilations and special conformal transformations \cite{NRT}
\footnote{Ref.~\cite{dRT} proposes a different realization of the conformal group on $\pi$.}.
At the infinitesimal level, the former are still linear in $x^\mu$, thus yielding conservation laws similar to those we discussed here. On the other hand, infinitesimal special conformal transformations are quadratic in $x^\mu$. Among the associated conservation laws, there will be one implying the existence of a local functional that grows {\em quadratically} with time.

\noindent
{\em Acknowledgements.}
I am grateful to K.~Hinterbichler, L.~Hui, and R.~Rattazzi for interesting discussions, and to E.~Trincherini for comments on the manuscript.
This work is supported by the DOE (DE-FG02-92-ER40699) and by NASA ATP
(09-ATP09-0049).


\end{document}